\documentclass[aps,pra,reprint,onecolumn,amssymb,showpacs]{revtex4}
\usepackage[paperwidth=210mm,paperheight=297mm,centering,hmargin=2cm,vmargin=2.5cm]{geometry}
\usepackage{standalone}

\usepackage{psfrag,graphicx}
\usepackage{tikz}
\usepackage{pgfplots}
\usepackage{dcolumn}
\usepackage{bm}
\usepackage{amsfonts,amssymb,amsmath} 
\usepackage{xcolor}
\usepackage{tikz}
\usetikzlibrary{decorations.markings}
\usetikzlibrary{arrows}
\usepackage{bbold}
\usepackage{epstopdf}
\usepackage{mathbbol}
\usepackage{amsfonts}
\usepackage{float}
\usepackage{subfig}

\usepackage{setspace}
\usepackage{enumerate}

\graphicspath{ {PhDimages/} }

\newcommand{\be}{\begin{equation}}
\newcommand{\ee}{\end{equation}}
\newcommand{\bq}{\begin{eqnarray}}
\newcommand{\eq}{\end{eqnarray}}
\newcommand{\ba}{\begin{align}}
\newcommand{\ea}{\end{align}}

\newcommand{\rf}[1]{(\ref{#1})}

\newcommand{\ket}[1]{\left |#1 \right\rangle}

\usepackage{tikz}

\usepackage{verbatim}

\usetikzlibrary{positioning,shapes.misc,calc,decorations.markings}
\tikzset{->-/.style={decoration={
			markings,
			mark=at position #1 with {\arrow{latex}}},postaction={decorate}}}
\tikzset{-<-/.style={decoration={
			markings,
			mark=at position #1 with {\arrow{latex reversed}}},postaction={decorate}}}
\usetikzlibrary{shapes.geometric}

\begin{document}

\title{Free fermion representation of the topological surface code}

\author{Ashk Farjami}
\affiliation{School of Physics and Astronomy, University of Leeds, Leeds, LS2 9JT, United Kingdom}

\date{\today}

\begin{abstract}
	The toric code is known to be equivalent to free fermions. This paper presents explicit local unitary transformations that map the $\mathbb{Z}_2$ toric and surface code --- the open boundary equivalent of the toric code --- to fermions. Through this construction it is shown that the surface code can be mapped to a set of free fermion modes, while the toric code requires additional fermionic symmetry operators. Finally, it is demonstrated how the anyonic statistics of these codes are encoded in the fermionic representations.
\end{abstract}
\maketitle


\section{Introduction}
The toric code \cite{KitaevToricCode,2and3DToricCodeReview,DanBrowneNotes} and its open boundary version, the $\mathbb{Z}_2$ surface code \cite{SurfaceVertex,BravyiMajor}, have been the test-bed for numerous investigations of condensed matter phenomena as well as quantum information applications \cite{SelfReview,SurfaceReview,ToricKitReview}. The main reasons for the popularity of the toric code are its ability to support Abelian anyons, exotic quasiparticles that can fault-tolerantly encode and manipulate quantum information, its eigenstates have non-trivial topological entanglement entropy \cite{HammaEntropy}, while it is exactly solvable. An important feature of this topological model is that it is relatively simple, where for example, the anyonic statistics and fusion rules emerge directly from the algebraic properties of Pauli matrices. At the same time the toric code enjoys many applications. It can be used as a fault tolerant quantum memory protecting against spurious local perturbations \cite{WoottonToricFT}, it can perform topological quantum computation resilient against control errors \cite{KitaevToricCode}, or it can encode more complex anyonic models such as Majorana fermions at lattice defects \cite{WoottonDefectMaj,BrownDefectMaj}.

The toric code has been experimentally simulated with highly entangled four-photon GHZ states \cite{photonTC} and the four-body interaction has been physically realised with Josephson junctions \cite{ToricJosephson,TopologJosephson}. However, it has been argued by Wen that the Hilbert space of the toric code, in the presence of an external magnetic field contains a low energy subspace that can be described effectively by hopping fermionic excitations coupled to a $\mathbb{Z}_2$ gauge field \cite{WenFreeArg}. This gauge field does not introduce interactions, but encodes the exotic statistics of the excitations. Moreover, previous investigation of the toric code's ground state in terms of the interaction distance \cite{DFNat} showed that the system is equivalent to free fermions \cite{DFPRB}. As this paper will show this is part of a more general result, where all eigenstates of the toric code are Gaussian states having entanglement spectra given in terms of free fermions. In addition, the energy spectrum has a similar decomposition in terms of single particle energies. Hence, it is expected that a unitary transformation exists that maps the toric code to a free fermion Hamiltonian. Nevertheless, a free fermion system can support neither anyonic statistics nor eigenstates with non-trivial topological entanglement entropy. Hence, these properties have to be encoded non-trivially in the unitary transformation that maps between these two physically different models.

Previous works studying transformations of the toric code include the paper \cite{BenFree2Toric}, where the authors provide a duality mapping from a cluster state on an $N\times N$ lattice to the toric code on an $N\times (N-1)$ lattice. The cluster state can be mapped to individual spin Hamiltonians, which are equivalent to free fermions. The mapping to the toric code takes some of the cluster state's boundary terms to stabilizers of non-contractible loops in the toric code, thus removing the degeneracy of the ground state. In addition, the paper \cite{ZoharMapping} maps the toric code onto decoupled Ising chains, and the papers \cite{CNOTToricToIsing,CNOTToricOpenToIsing} give duality mappings, built from CNOT gates, from the toric and surface code in the presence of external magnetic fields to Ising models.

This paper demonstrates that indeed it is possible to find a unitary transformation that maps the surface code to free fermions and presents its explicit form. It also presents the explicit form of a unitary transformation mapping the toric code to free fermions with an interacting fermionic parity operator, which ensures the excitations of the model are created in pairs, as in the toric code. These transformations comprise of products of $C_4$ Clifford rotations \cite{C4rot} that act on each plaquette, and are directly generalisable to arbitrary size systems. The resulting system of the surface code transformation consists of free fermion modes with local chemical potentials, that can encode the excitations of the plaquettes, and of a single zero energy fermion mode that does not appear in the Hamiltonian, encoding the logical state of the model. The toric code supports quasiparticle excitations that come always in pairs, while the surface code can have individual excitations. On the other hand the free fermion system described can support single particle excitations. As the unitary transformation is isospectral, it cannot map the toric code to a system of this form. The toric code is mapped to a similar system, with one extra zero mode to encode the second logical qubit of the toric code and a fermionic parity operator ensuring any excitations are created in pairs, thus fixing the isospectral nature of the transformation. The possibility to transform the surface code to free fermions could have a variety of applications, e.g. in condensed matter, by dissecting the way anyonic statistics emerge, or in quantum information, as free fermion systems and their manipulation have a very efficient descriptions \cite{BravyiMajor,BravyiFLO}.

The paper is organised as follows. Section II A reviews the spin description of the surface and toric code, which will be the starting models. Section II B explicitly presents the local unitary transformations, $\mathcal{U}_S$ and $\mathcal{U}_T$, that map between the surface and toric code and their fermionic counterparts, respectively, for arbitrary size systems. Section II C studies the resulting models, showing how the states of the models split into ``dynamic" and ``zero" (or ``logical") modes. Section III looks at how string operators transform between the systems. It is demonstrated how the mapping keeps endpoints of anyonic string operators fixed, while extending their support into the dynamic and logical modes of the fermionic systems. Finally, It is shown how the anyonic statistics of the surface code are encoded in excitations of the free fermion model and how the same statistics of the toric code are encoded in its fermionic counterpart.


\section{The transformation to fermions}

In order to perform the transformation of the $\mathbb{Z}_2$ surface \cite{BravyiMajor} and toric code \cite{KitaevToricCode} to free fermions and fermions with fermionic parity operators, respectively, let us first adopt a suitable representation of the models in terms of spins located at the vertices of the lattice \cite{SurfaceVertex}. We will then study the energy and entanglement spectra and show that they both exactly correspond to those of free fermions. This description can be formalised through the use of the interaction distance, $D_{\mathcal{F}}$, \cite{DFNat,DFPRB}. We then see that the stabilizer groups of both the surface and toric codes are isomorphic to groups generated by commuting Pauli operators, thus giving a description of the form of these unitarily equivalent fermion models. Two explicit unitaries, $\mathcal{U}_S$ and $\mathcal{U}_T$, are then presented. The first of which transforms the surface code to decoupled fermions, and the second transforms the toric code to fermions coupled to two fermionic parity operators.


\subsection{The spin description of the surface and toric code}
\begin{figure}[t]
	\includegraphics[width = 0.85\linewidth]{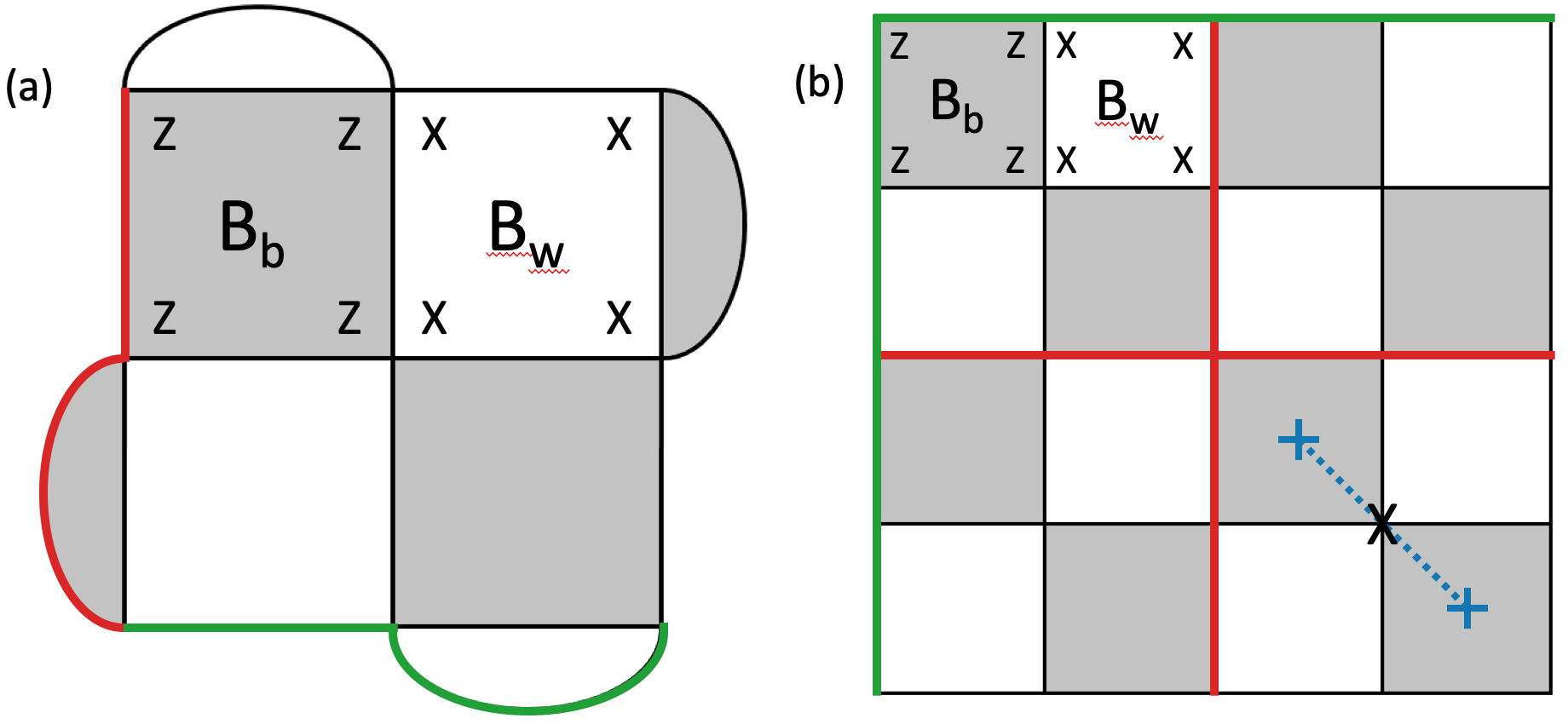}
	\caption{(a) The $3\times3$ surface code, with physical qubits located at the vertices of the lattice. Plaquette stabilizers, $B_b$ and $B_w$, are on black, $b$, and white, $w$, plaquettes respectively. $B_b$ ($B_w$) applies a $\sigma^z$ ($\sigma^x$) operator to each qubit surrounding $b$ ($w$) and detects the parity of $\sigma^x$ ($\sigma^z$) operators on these qubits. Logical Pauli operators $X_L$ and $Z_L$ are shown as the red and green lines, respectively, encoding one logical qubit. (b) The $4\times4$ toric code, with physical qubits at the vertices of the lattice. Plaquette stabilizers, $B_b$ and $B_w$, are of the same form as those in the surface code. The choices of $X$ and $Z$ logical operators are shown in red and green, respectively, with $X_L^{(1)}$ and $Z_L^{(2)}$ depicted as the horizontal lines and $Z_L^{(1)}$ and $X_L^{(2)}$ as the vertical lines. These encode the two logical qubits of the toric code. A $\sigma^x$ operator acting on a single qubit producing a string with an excitation on the plaquettes at each end, is shown in blue.}
	\label{fig:intsurfaceandtoric}
\end{figure}

We now review the spin description of the surface and toric code \cite{SurfaceVertex}. We start with the surface code. The distance $d$ surface code has support on a $d\times d$ lattice with open boundary conditions, where $d$ is always odd. There are $d^2$ physical qubits, or spins, one on each site of the lattice, that encode one logical qubit. The code has an alternating checker pattern of stabilizers, $B_b$ and $B_w$, at the black, $b$, and white, $w$, plaquettes, built from $\sigma^x$ or $\sigma^z$ Pauli operators on the surrounding physical qubits, respectively, as shown in Fig.~\ref{fig:intsurfaceandtoric}(a). The stabilizers corresponding to the semicircle plaquettes on the boundary and square plaquettes have support on two and four qubits, respectively. We can define two logical operators, $X_L$ and $Z_L$, as strings of $\sigma^x$ and $\sigma^z$ operators, respectively, with support on the $d$ physical qubits along the left and bottom of the lattice, respectively. These map between degenerate states of the model and are built from $\sigma^x$ and $\sigma^z$ operators, respectively. The Hamiltonian is given by
\be
H_{\text{SC}}=-J\sum_bB_b -J\sum_wB_w
\label{eq:HamSC}
\ee
where $B_b=\prod_{j\in b}\sigma^z_j$ and $B_w=\prod_{j\in w}\sigma^x_j$, with $\{b\}$ and $\{w\}$ the set of all black and white plaquettes, respectively, and $j$ is the site the Pauli operator acts on. The form of the Hamiltonian, $H_{\text{SC}}$, suggests that it is strongly interacting. Excitations in this model arise due to string operators, $O_C^X$ and $O_C^Z$, which are strings of $\sigma^x$ and $\sigma^z$ operators respectively,
\be
\begin{aligned}
	O_C^X=\prod_{j \in C}\sigma^x_j, \;\;\;\;\; O_C^Z=\prod_{j \in C}\sigma^z_j,
\end{aligned}
\label{eq:stringint}
\ee
that create localised excitations at the endpoints of the path, $C$, of the string operator. $O_C^X$ operators create excitations at black plaquettes and $O_C^Z$ at white plaquettes. Crossings between these string operators on qubits give rise to the anyonic statistics through the Pauli commutation relations. A single $\sigma^x$ or $\sigma^z$ is the smallest string operator, having endpoints on two plaquettes, diagonally adjacent to each other. This is the standard form of the surface code in terms of qubits (spins).

The toric code is similar to the surface code. The distance $d$ toric code has support on a $d\times d$ lattice on the surface of a torus with periodic boundary conditions, where $d$ is always even. It still has $d^2$ physical qubits, but now encodes two logical qubits and supports four logical operators, $X_L^{(1)}$, $Z_L^{(1)}$, $X_L^{(2)}$ and $Z_L^{(2)}$. These operators can be chosen arbitrarily as long as $X_L^{(1)}$ and $Z_L^{(2)}$ loop around the same non-contractible loop of the torus, $Z_L^{(1)}$ and $X_L^{(2)}$ loop around the other, and the following commutation and anti-commutation relations are obeyed, $[X_L^{(1)},X_L^{(2)}]=[Z_L^{(1)},Z_L^{(2)}]=[X_L^{(1)},Z_L^{(2)}]=[Z_L^{(1)},X_L^{(2)}]=0=\{X_L^{(1)},Z_L^{(1)}\}=\{X_L^{(2)},Z_L^{(2)}\}$.  For any distance $d$ toric code we choose all logical operators to be of length $d$ and parallel loops ($X_L^{(\gamma)}$ and $Z_L^{(\tau)}$, where $\gamma\neq\tau$) to be a distance $d/2$ from each other. This choice is shown for the $4\times4$ toric code in Fig.~\ref{fig:intsurfaceandtoric}(b). The code has an alternating checker pattern of square plaquettes, with support on four qubits each. Most other aspects of the two codes are equivalent, including the Hamiltonian,
\be
H_{\text{TC}}=-J\sum_bB_b -J\sum_wB_w.
\label{eq:HamTC}
\ee
Excitations of the toric code are produced by string operators of the same form as those in the surface code, given in \rf{eq:stringint}. A string operator on the toric code, consisting of a single $\sigma^x$, is shown in Fig.~\ref{fig:intsurfaceandtoric}(b). One consequence of the periodic boundary conditions is that $\prod_{b}B_b=\prod_{w}B_w=\mathbb{1}$, this will prove to be important when producing the unitary mapping from the toric code to free fermions with fermionic parity operators.

To investigate how close to free fermions these codes are, we can employ the interaction distance. This distance is the minimal trace distance of a given density matrix to the manifold, $\mathcal{F}$, of all Gaussian density matrices,
\be
D_{\mathcal{F}}(\rho)=\min_{\sigma\in\mathcal{F}}D(\rho,\sigma),
\label{eq:IntDist}
\ee
where $D(\rho,\sigma)=\frac{1}{2}\text{tr}\sqrt{(\rho-\sigma)^2}$ is the trace distance. A more convenient expression for the interaction distance is in terms of the eigenstates of $\rho$ and $\sigma$, i.e. $D_{\mathcal{F}}(\rho)=\min_{\{\sigma_k\}\in\mathcal{F}}\sum|\rho_k-\sigma_k|$. The interaction distance can be determined for $\rho$ being the reduced density matrix from the bipartition of the system's ground state or being the thermal density matrix. In the first case $D_{\mathcal{F}}$ probes the quantum correlations of the system and thus its entanglement spectrum, while in the second case it probes its energy spectrum. Recent studies of the interaction distance, $D_{\mathcal{F}}$, \cite{DFNat,DFPRB}, of the $\mathbb{Z}_2$ surface and toric code have found that $D_{\mathcal{F}}=0$, for the entanglement spectrum of the ground state. Here we show that not only does the ground state entanglement spectrum correspond to that of free fermions but so does the energy spectrum. Let us first consider in more detail the interaction distance with respect to the entanglement and energy spectra of the model. The entanglement interaction distance, $D^{\text{ent}}=D_{\mathcal{F}}(\rho^{\text{ent}})$, is the interaction distance of the reduced density matrix, $\rho^{\text{ent}}$, produced by a given bi-partition of the model in a particular eigenstate. In general, we can write the eigenvalues of $\rho^{\text{ent}}$ as $\rho_k^{\text{ent}}=e^{-E_k^{\text{ent}}}$, where $\{E_k^{\text{ent}}\}$ is the entanglement spectrum of the state. For Gaussian states the entanglement spectra are given by,
\be
E_k^{\text{free}}=E_0+\sum_{j=1}^{N}\epsilon_jn_j(k), \; k=1,2,...,2^N,
\label{eq:FreeElvls}
\ee
where $\epsilon_j$ are the single particle energies of the free fermion modes, $n_j(k)$ are the single particle occupations corresponding to the energy level $k$ and $N$ is the number of fermion modes in the system \cite{DFJandZ}. Bi-partitioning the state of an anyonic system into two parts $A$ and $B$ gives an entanglement spectrum of the form,
\be
E_k^{\text{ent}}=-\log(\frac{\mathcal{N}^c_{\boldsymbol{a}}\prod_{l\in \boldsymbol{i}}d_{a_l}}{\mathcal{D}^{|\partial A|-1}}),
\ee
where $\boldsymbol{i}$ is a specific anyonic configuration related to $k$, $\boldsymbol{a}$ is the set of all anyons in $\boldsymbol{i}$, $c$ is the total anyonic charge across the bi-partition, $\mathcal{N}$ is the multiplicity of the fusion of the anyons $\boldsymbol{a}$ to $c$, $d_{a_l}$ is the quantum dimension of an anyon $a_l$, $\mathcal{D}$ is the total quantum dimension of the system and $|\partial A|$ is the size of the boundary of $A$ \cite{Alex,Levin}. Then $|\alpha_{\boldsymbol{i}}|^2=\frac{\mathcal{N}^c_{\boldsymbol{a}}\prod_{l\in \boldsymbol{i}}d_{a_l}}{\mathcal{D}^{|\partial A|-1}}$ is the normalised probability of having an anyonic configuration, $\boldsymbol{i}$, at the boundary of the partition, with total charge $c$. The entire spectrum is built from all possible anyonic configurations, $\boldsymbol{i}$. The surface and toric code are abelian models, so $\mathcal{N}_{\boldsymbol{a}}^c=1$ for all valid sets of anyons $\boldsymbol{a}$ that fuse to $c$ and zero otherwise. The quantum dimensions, $d_{a_l}$, of all anyons of the these codes are equal to one and there are four species of anyon \cite{ToricKitReview}. Hence, $\mathcal{D}=\sqrt{\sum_ad_a^2}=2$, where the sum is over all anyons of the code. Therefore each state of the surface and toric code has a flat entanglement spectrum with degeneracy proportional to the size of the boundary of the partition $E_k^{\text{ent}}=-\log(\frac{1}{2^{|\partial A|-1}})$, \cite{DFPRB,Alex}. This spectrum has the same form as that of (\ref{eq:FreeElvls}) with all $\epsilon_j$'s set to zero. Therefore $D^{\text{ent}}=0$ for the entanglement spectrum of all states of the surface and toric code for all possible partitions.

The thermal interaction distance, $D_{\text{th}}^{\beta}=D_{\mathcal{F}}(\rho^{\text{th}}(\beta))$, is the interaction distance of the thermal density matrix, $\rho^{\text{th}}=\frac{1}{Z}e^{-\beta H}$, where $Z=\text{tr}(e^{-\beta H})$ is the partition function and $T=\frac{1}{\beta}$ is the temperature. The eigenvalues of $\rho^{\text{th}}$ have the form $\rho_k^{\text{th}}=\frac{1}{Z}e^{-\beta E_k}$, where $\{E_k\}$ is the energy spectrum. The energy spectrum of free states should satisfy the same relation as (\ref{eq:FreeElvls}). For the surface and toric code, the energy spectrum, $E_k$, is given by the syndrome pattern of anyonic excitations at plaquettes. These excitations all have the same energy contribution, as seen from \rf{eq:HamSC} and \rf{eq:HamTC}, hence the spectrum of the surface code can be reproduced with a set of $d^2-1$ single particle energies, $\{\epsilon_j\}$, (corresponding to the $d^2-1$ plaquettes of the $d\times d$ surface code) arranged in all possible occupation patterns, $n_j(k)$. The spectrum of the toric code can be reproduced with a set of $d^2$ single particle energies, (corresponding to the $d^2$ plaquettes of the $d\times d$ toric code) arranged in all possible occupation patterns, with even total occupation number. Therefore, the thermal interaction distance of these codes is zero. The fact that $D_{\text{th}}^{\beta}=0$ means these codes are isospectral to free fermion systems. This suggests, there should exist unitary transformations, $\cal{U}$, mapping the surface and toric code presented above to such isospectral free fermion models.

By studying the group structure that corresponds to the surface and toric code we find the form of the fermionic models they map to. The group generated by the set of all surface code stabilizer operators, $\{B_p\}_{\in \mathcal{P}}$, where $\mathcal{P}$ is the set of all plaquettes, all stabilizers square to the identity and stabilizers commute with one another, is isomorphic to a group generated by a set of commuting Pauli operators.
\be
\left\langle B_p|B_p^2=B_{\lambda}B_{\eta}B_{\lambda}B_{\eta}=\mathbb{1}\right\rangle\cong\left\langle\sigma^z_{i\forall i=1,...,|\mathcal{P}|}\right\rangle.
\label{eq:SurfaceIsom}
\ee
This suggests it should be possible to map each plaquette operator of the surface code to a single free fermion mode.

The toric code stabilizer group has the added restriction that the product of all stabilizers supported on a black, $B_b$, or white, $B_w$, plaquette, respectively, must be equal to the identity. The group generated by these stabilizer operators is isomorphic to a group generated by a set of commuting Pauli operators two smaller than that of the surface code,
\be
\left\langle B_p|B_p^2=B_{\lambda}B_{\eta}B_{\lambda}B_{\eta}=\prod_{p\in \mathcal{P}_b}B_p=\prod_{p\in \mathcal{P}_w}B_p=\mathbb{1}\right\rangle\cong\frac{\left\langle B_p|B_p^2=B_{\lambda}B_{\eta}B_{\lambda}B_{\eta}=\mathbb{1}\right\rangle}{\left\langle\prod_{p\in \mathcal{P}_b}B_p=\prod_{p\in \mathcal{P}_w}B_p=\mathbb{1}\right\rangle}\cong\left\langle\sigma^z_{i\forall i=1,...,|\mathcal{P}|-2}\right\rangle.
\label{eq:ToricIsom}
\ee
The resulting group generated by the Pauli operators in (\ref{eq:ToricIsom}) will be one quarter the size of that in (\ref{eq:SurfaceIsom}). All plaquette operators in the toric code should be mapped to free fermion modes, except one black and one white operator which will each be mapped to fermionic parity operators over the set of all black and white modes respectively. Hence, even though the interaction distance tells us that the toric code is isospectral to a free fermion model, we actually find these interacting fermionic parity operators are necessary by studying the group structure. These symmetry terms are a result of the periodic boundary conditions of the toric code and the fact that excitations in the code are created in pairs. This is discussed in more detail in section III.

The surface and toric code are mapped by unitary transformations, $\mathcal{U}_S$ and $\mathcal{U}_T$, respectively, to fermionic models
\be
\begin{aligned}
	{\cal U}_S H_{\text{SC}}{\cal U}_S^{\dagger} &= H_{\text{FS}}\\
	{\cal U}_T H_{\text{TC}}{\cal U}_T^{\dagger} &= H_{\text{FT}}
\end{aligned}
\label{eq:HamMap}
\ee
where $H_{\text{FS}}$ is a free fermion model Hamiltonian and $H_{\text{FT}}$ is a fermionic model Hamiltonian, consisting of free fermion terms and two interacting fermionic parity operators. It is the purpose of the next section to present the exact form of $\mathcal{U}_S$ and $\mathcal{U}_T$.


\subsection{The unitary transformations}

This section presents the transformations, $\mathcal{U}_S$ and $\mathcal{U}_T$, between the spin representation of the surface and toric code and the fermionic Hamiltonians, $H_{\text{FS}}$ and $H_{\text{FT}}$, respectively, as dictated by (\ref{eq:HamMap}). These are general unitaries for any system size. To achieve this we employ $C_4$ Clifford rotations \cite{C4rot}, of the form,
\be
R_{C_4}(\sigma^{[\mu]})=\text{exp}(\frac{i}{\pi}\sigma^{[\mu]})=\frac{1}{\sqrt{2}}(1+i\sigma^{[\mu]})
\ee
where $\sigma^{\mu_i}$ is the Pauli matrix acting on the $i$th qubit and $\sigma^{[\mu]}\equiv\sigma^{\mu_1\mu_2\mu_3...}\equiv\sigma^{\mu_1}\otimes\sigma^{\mu_2}\otimes\sigma^{\mu_3}\otimes...$ is the direct product of some set of Pauli matrices. The action of $R_{C_4}$ on a matrix $\sigma^{[\nu]}$ is given by,
\be
\begin{aligned}
	\sigma^{[\nu]}&\rightarrow R_{C_4}^{\dagger}(\sigma^{[\mu]})\sigma^{[\nu]}R_{C_4}(\sigma^{[\mu]})\\
	&=\begin{cases}
		\sigma^{[\nu]}, &\text{ if } [\sigma^{[\nu]},\sigma^{[\mu]}]=0,\\
		i\sigma^{[\nu]}\sigma^{[\mu]}, &\text{ if } \{\sigma^{[\nu]},\sigma^{[\mu]}\}=0.
	\end{cases}
\end{aligned}
\ee
Using this wet can map a collection of spin operators to a spin operator on a single qubit and the identity everywhere else. For example,
\be
\begin{aligned}
	\sigma^{[\mu]x[\nu]}&\xrightarrow{R_{C_4}(-\sigma^{[\mu]y[\nu]})}-i\sigma^{[\mu]x[\nu]}\sigma^{[\mu]y[\nu]}=\sigma^{[0...]z[0...]}\\
	\sigma^{[\mu]z[\nu]}&\xrightarrow{R_{C_4}(\sigma^{[\mu]y[\nu]})}i\sigma^{[\mu]z[\nu]}\sigma^{[\mu]y[\nu]}=\sigma^{[0...]x[0...]},
\end{aligned}
\ee
where $\sigma^0=\mathbb{1}$.

\begin{figure}[t]
	\includegraphics[width = 0.9\linewidth]{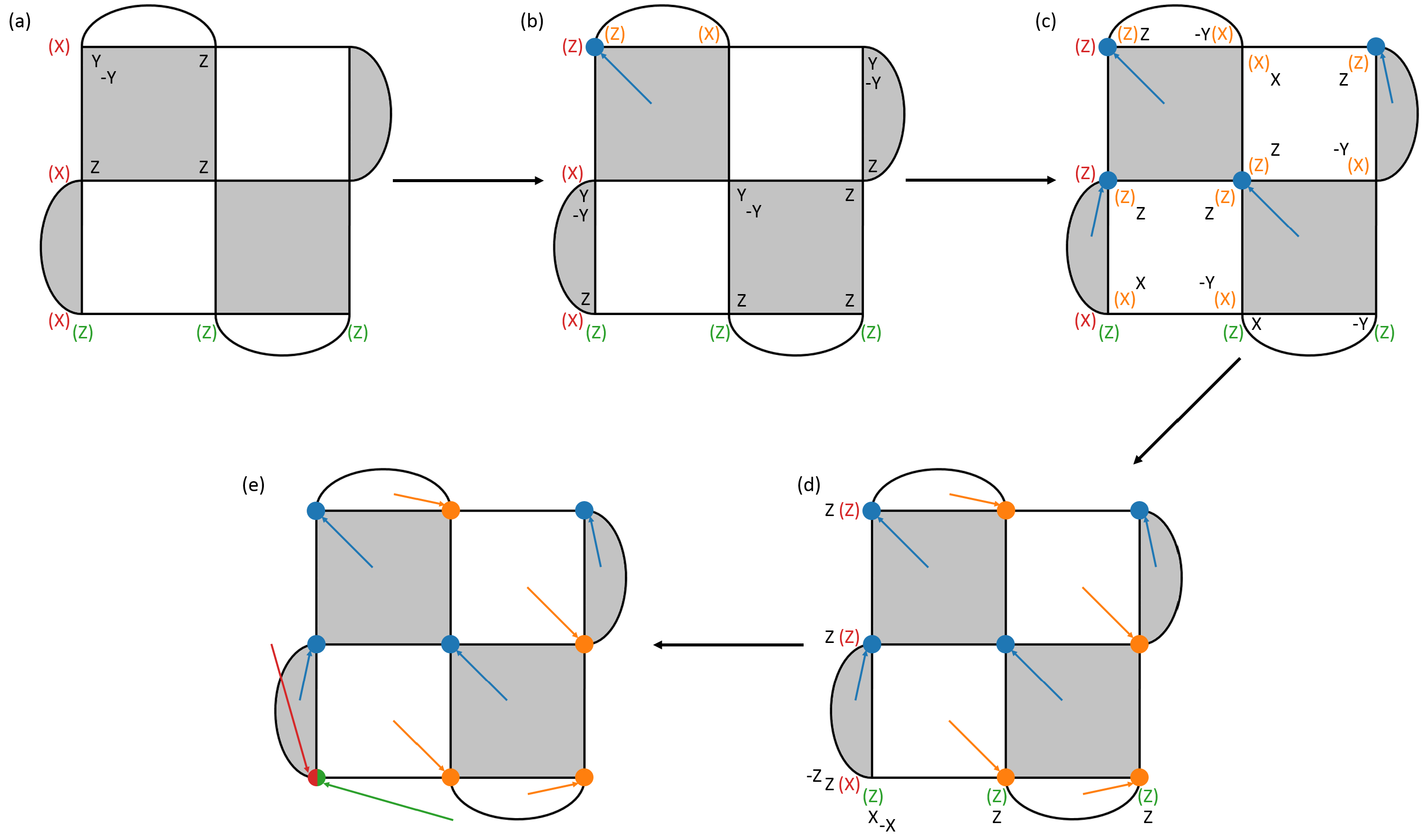}
	\caption{Mapping of the $3\times3$ surface code under $\mathcal{U}_S$. (a) Rotations $R_{C_4}(\sigma^{[0...]yzzz[0...]})$ and $R_{C_4}(-\sigma^{[0...]y000[0...]})$ acting on the top left black plaquette stabilizer, $B_{b_1}$, are labelled in black on the interior of $b_1$. (b) Blue arrows indicate the free fermion modes supporting transformed black plaquette stabilizers, $\tilde{B}_{b}$. The letters in parentheses show the form of operators acted on non-trivially by the rotations. The letters are orange for white plaquettes, red for $X_L$ and green for $Z_L$. Rotations for the other three black plaquettes are labelled in black in their interior. (c) The $C_4$ rotations corresponding to the white plaquettes are shown labelled in black in their interior. (d) Orange arrows show the positions of free fermion modes supporting transformed white plaquette stabilizers, $\tilde{B}_{w}$. The rotations mapping $(U_N...U_1)X_L(U_1^{\dagger}...U_N^{\dagger})$ and $(U_N...U_1)Z_L(U_1^{\dagger}...U_N^{\dagger})$ to a single $\sigma^x$ and $\sigma^z$, respectively, with support on the logical mode are labelled along the left and bottom of the lattice. (e) Red and green arrows point to the logical mode supporting $\tilde{X}_L$ and $\tilde{Z}_L$.}
	\label{fig:planarmap}
\end{figure}

Let us start with the surface code mapping, $\mathcal{U}_S$. The purpose of $\mathcal{U}_S$ is to transform each plaquette stabilizer, $B_p$, in $H_{\text{SC}}$ to an operator, $\tilde{B}_p=\sigma^z=1-2a^{\dagger}a$ \cite{NielsenFermiSpinNotes} ($a^{\dagger}$ and $a$ are fermionic creation and annihilation operators, respectively), on a single free fermion mode (or spin) and the logical operators, $X_L$ and $Z_L$, to operators $\tilde{X}_L=\sigma^x=a^{\dagger}+a$ and $\tilde{Z}_L=\sigma^z=1-2a^{\dagger}a$ with support on a shared zero mode, not in the Hamiltonian, $H_{\text{FS}}$, hence separate from those supporting $\tilde{B}_p$ operators. We split $\mathcal{U}_{S}$ into $N+2$ unitaries,
\be
\mathcal{U}_{S}=U_{N+2}...U_2U_1,
\label{eq:decomposeP}
\ee
where $N=d^2-1$ is the number of plaquettes in the $d\times d$ surface code. Each of the $U$'s has a similar structure, transforming one of the $d^2-1$ plaquette stabilizers, or $2$ logical operators into single spin operators. The first $N/2$ unitary parts, $\{U_1,...,U_{N/2}\}$, correspond to the transformation of black ($Z$) plaquette operators with support on four qubits, $B_b^{(4)}=\sigma^{[0...]zzzz[0...]}$, or two qubits, $B_b^{(2)}=\sigma^{[0...]zz[0...]}$, as shown in Fig.~\ref{fig:planarmap}(a) and (b). The mappings take the following form,
\be
\begin{aligned}
	B_b^{(4)}&\xrightarrow{R_{C_4}(\sigma^{[0...]yzzz[0...]})}\sigma^{[0...]x000[0...]}\xrightarrow{R_{C_4}(-\sigma^{[0...]y000[0...]})}\sigma^{[0...]z000[0...]},\\
	B_b^{(2)}&\xrightarrow{R_{C_4}(\sigma^{[0...]yz[0...]})}\sigma^{[0...]x0[0...]}\xrightarrow{R_{C_4}(-\sigma^{[0...]y0[0...]})}\sigma^{[0...]z0[0...]},
\end{aligned}
\label{eq:UblackP}
\ee
where $R_{C_4}(\sigma^{[0...]yzzz[0...]})R_{C_4}(-\sigma^{[0...]y000[0...]})$ is one unitary part, $U$. We label the resulting operators $\tilde{B}_b^{(4)}=\sigma^{[0...]z000[0...]}$ and $\tilde{B}_b^{(2)}=\sigma^{[0...]z0[0...]}$. These operators have support on the top left qubit of the corresponding plaquette, $b$, or just the top qubit for the order two stabilizers. All other Pauli operators in the $C_4$ rotation are equal to those in the operator we are mapping from at each stage, but at this top left qubit we replace the Pauli operator with a $\sigma^y$ in the first step and a $-\sigma^y$ in the second. There is a lot of freedom in the choice of the specific form of $C_4$ rotations throughout this section. For example, we could have a $-\sigma^y$ in the first $R_{C_4}$ and a $\sigma^y$ in the second. We just present two particular collections of $C_4$ rotations that work for the surface and toric code, respectively. These $N/2$ unitary parts act in order from the top to bottom row of the lattice. This ensures that their effect on all other black plaquettes are trivial. The effect of these unitaries on the white ($X$) plaquettes, however are non-trivial. We see in Fig.~\ref{fig:planarmap}(b) and (c) that some $\sigma^x$ operators of the white plaquettes are mapped to $\sigma^z$ operators by the first $N/2$ unitaries. These $\sigma^x$ operators are those with support on the qubits pointed at by the blue arrows.

The white ($X$) plaquette stabilizers are mapped, by the next $N/2$ unitary parts, to $\sigma^z$ operators on the bottom right qubit of the plaquette, or the right qubit in the case of the order two operators. They are mapped individually, each by their own $U$ in order from the right to left column. The form of the $U$'s that perform this mapping vary depending on the effect of the $U$'s corresponding to the black plaquettes. The mapping $(U_{N/2},...,U_1)$ acts trivially on the semi circle plaquette stabilizers on the bottom row of the lattice, as is shown in Fig.~\ref{fig:planarmap}(c), $(U_{N/2}...U_1)B_w^{(2)}(U_1^{\dagger}...U_{N/2}^{\dagger})=B_w^{(2)}=\sigma^{[0...]xx[0...]}$. For these types of plaquettes we use the rotation,
\be
B_w^{(2)}\xrightarrow{R_{C_4}(-\sigma^{[0...]xy[0...]})}\sigma^{[0...]0z[0...]}
\ee
All others are acted on non-trivially, such as the top right square plaquette stabilizer, $B_w^{(4)}$, in Fig.~\ref{fig:planarmap}(c), $(U_{N/2}...U_1)B_w^{(4)}(U_1^{\dagger}...U_{N/2}^{\dagger})=\bar{B}_w^{(4)}=\sigma^{[0...]xzzx[0...]}$, where $\bar{B}_w^{(4)}$ labels the intermediate form of the operator. For an operator of this form we use the rotation,
\be
\bar{B}_w^{(4)}\xrightarrow{R_{C_4}(-\sigma^{[0...]xzzy[0...]})}\sigma^{[0...]000z[0...]},
\ee
where the Pauli operators in the $R_{C_4}$ rotation are equal to those in the operator we are mapping from, $\bar{B}_w^{(4)}$, except at the bottom right qubit of the plaquette where we replace the $\sigma^x$ with a $-\sigma^y$. The operator on the bottom right qubit of a white plaquette is always unaffected by any previous $U$'s by construction, thus will remain a $\sigma^x$.

Once we have transformed the $N$ plaquette operators, we transform the logical operators with the two remaining unitaries, $U_{N+1}$ and $U_{N+2}$. The logical operator, $X_L$ is mapped by all previous unitaries to a string of $\sigma^z$ operators along the left boundary attached to a $\sigma^x$ on the bottom left qubit of the lattice, where it intersects with $Z_L$. At the same time $Z_L$ is acted on trivially by all previous unitaries. We label these intermediate forms of the operators as $\bar{X}_L=\sigma^{[0...]xzz...[0...]}$ and $\bar{Z}_L=\sigma^{[0...]zzz...[0...]}$, respectively. They are both shown in Fig.~\ref{fig:planarmap}(d), along with the form of $U_{N+1}$ and $U_{N+2}$, for a $3\times3$ lattice. These act as,
\be
\begin{aligned}
	\bar{X}_L&\xrightarrow{R_{C_4}(\sigma^{[0...]zzz...[0...]})}\sigma^{[0...]y00...[0...]}\xrightarrow{R_{C_4}(-\sigma^{[0...]z00...[0...]})}\sigma^{[0...]x00...[0...]},\\
	\bar{Z}_L&\xrightarrow{R_{C_4}(\sigma^{[0...]xzz...[0...]})}-\sigma^{[0...]y00...[0...]}\xrightarrow{R_{C_4}(-\sigma^{[0...]x00...[0...]})}\sigma^{[0...]z00...[0...]}
\end{aligned}
\ee
for a general size code. Thus $X_L$ and $Z_L$ are mapped to $\tilde{X}_L=\sigma^{[0...]x00...[0...]}$ and $\tilde{Z}_L=\sigma^{[0...]z00...[0...]}$, respectively, with support on a single shared qubit. $U_{N+1}$ and $U_{N+2}$ act trivially on all previously obtained $\tilde{B}_b$ and $\tilde{B}_w$ operators.

\begin{figure}[t]
	\includegraphics[width = 1.0\linewidth]{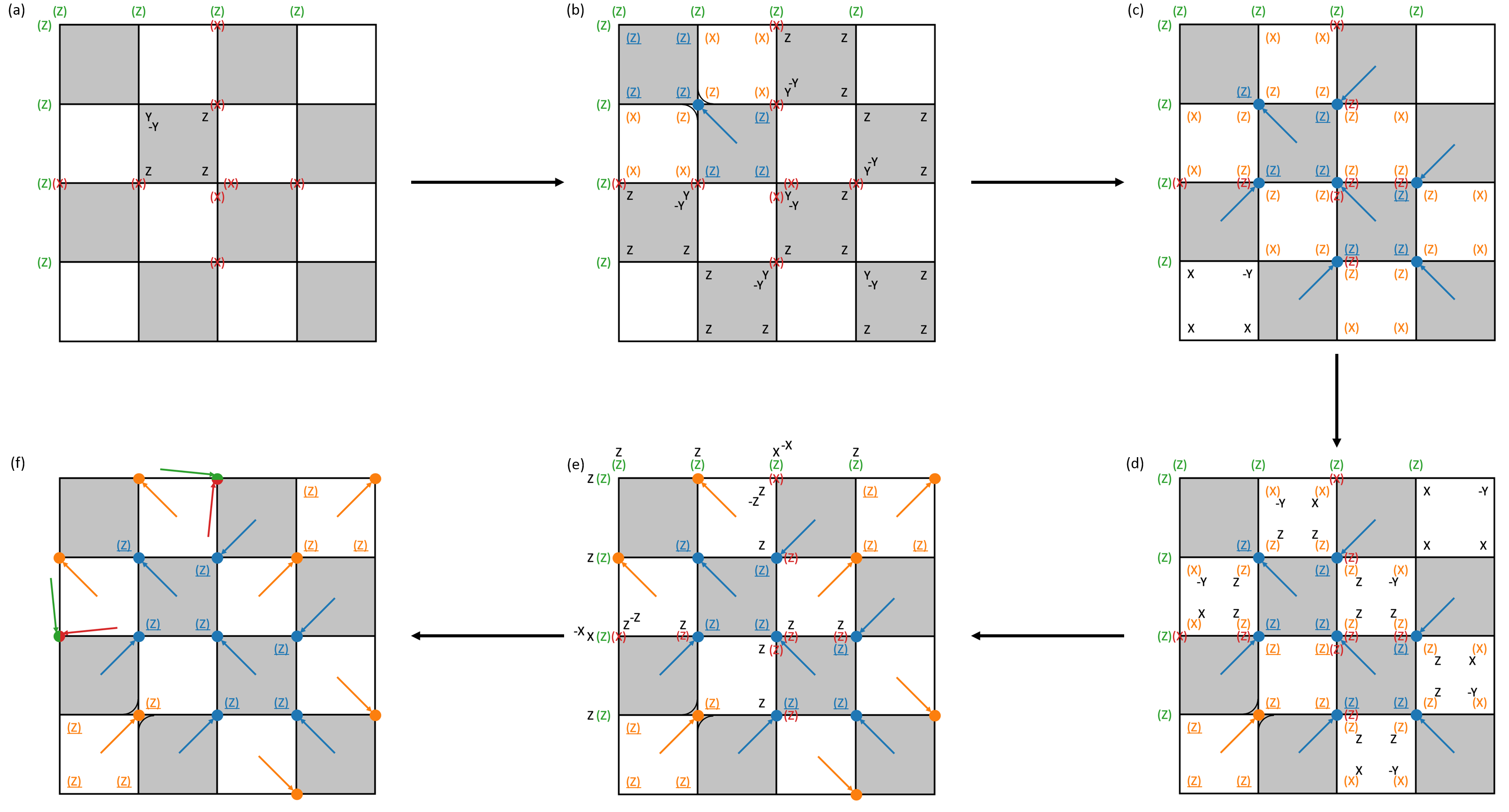}
	\caption{Mapping of the $4\times4$ toric code under $\mathcal{U}_T$ (a) Rotations $R_{C_4}(\sigma^{[0...]yzzz[0...]})$ and $R_{C_4}(-\sigma^{[0...]y000[0...]})$ acting on the black plaquette stabilizer, $B_{b_2}$, are labelled in black on the interior of $b_2$. (b) Blue arrows indicate the fermion modes supporting transformed black plaquette stabilizers, $\tilde{B}_{b}$. The letters in parentheses show the form of operators acted on non-trivially by the rotations. The letters are orange for white plaquettes, blue for black plaquettes, red for $X_L$ and green for $Z_L$. They are underlined for the two plaquette operators that will be mapped to fermionic parity operators. Rotations for the other black plaquettes are labelled in black in their interior. (c) The $C_4$ rotation corresponding to the bottom left white plaquette is shown labelled in black in its interior. (d) Rotations for the remaining white plaquettes are shown in their interior. Orange arrows show the positions of fermion modes supporting transformed white plaquette stabilizers, $\tilde{B}_{w}$. (e) Rotations mapping the partly transformed logical operators to single Pauli operators, with support the two logical modes are labelled along the paths of the original logical operators. (d) Red and green arrows point to the logical modes supporting $\tilde{X}_L^{(1)}$, $\tilde{Z}_L^{(1)}$, $\tilde{X}_L^{(2)}$ and $\tilde{Z}_L^{(2)}$. The black and white symmetry operators are labelled in blue and orange, respectively, with support on all transformed operators of the same colour.}
	\label{fig:toricmap}
\end{figure}

\begin{figure}[t]
	\includegraphics[width = 0.45\linewidth]{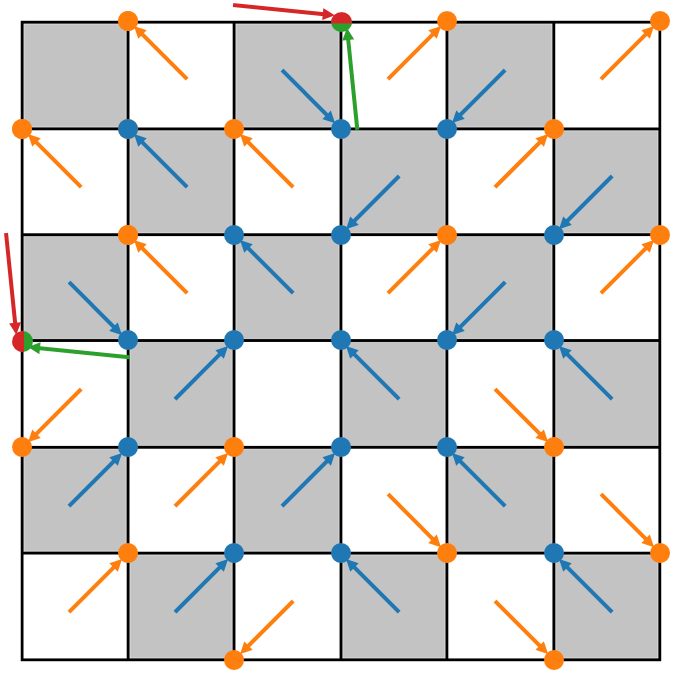}
	\caption{The orientation of all unitary parts, $U$, for the $6\times6$ toric code.}
	\label{fig:toriclarge}
\end{figure}

The toric code mapping, $\mathcal{U}_T$, has a similar form. $\mathcal{U}_T$ transforms each plaquette stabilizer, $B_p$, in $H_{\text{TC}}$ to an operator $\tilde{B}_p=\sigma^z$ on a single free fermion mode, except one black, $B_{b_1}$, and one white, $B_{w_1}$, stabilizer, which are mapped to symmetry operators, $\tilde{S}_{b_1}=\prod_{b\setminus b_1}\tilde{B}_b$ and $\tilde{S}_{w_1}=\prod_{w\setminus w_1}\tilde{B}_w$, which are the products of all other black and white transformed stabilizers respectively. The four logical operators, $X_L^{(1)}$, $Z_L^{(1)}$, $X_L^{(2)}$ and $Z_L^{(2)}$, are mapped to operators $\tilde{X}_L^{(1)}=\sigma^x_j$, $\tilde{Z}_L^{(1)}=\sigma^z_j$, $\tilde{X}_L^{(2)}=\sigma^x_k$ and $\tilde{Z}_L^{(2)}=\sigma^z_k$ with support on two zero modes, $j$ and $k$, not in the Hamiltonian, $H_{TS}$, hence separate from those supporting $\tilde{B}_p$ operators. We split $\mathcal{U}_{T}$ into $M+2$ unitaries,
\be
\mathcal{U}_{T}=U_{M+2}...U_2U_1,
\label{eq:decomposeT}
\ee
where $M=d^2$ is the number of plaquettes in the $d\times d$ toric code. Each of the $U$'s transforms one of the $d^2$ plaquette stabilizers, or $4$ logical operators into single spin operators. The first $M/2-1$ unitary parts, $\{U_1,...,U_{M/2-1}\}$, correspond to the transformation of black ($Z$) plaquette operators, $B_b=\sigma^{[0...]zzzz[0...]}$, as shown in Fig.~\ref{fig:toricmap}(a) and (b). The mappings take the following form,
\be
\begin{aligned}
	B_b&\xrightarrow{R_{C_4}(\sigma^{[0...]yzzz[0...]})}\sigma^{[0...]x000[0...]}\xrightarrow{R_{C_4}(-\sigma^{[0...]y000[0...]})}\sigma^{[0...]z000[0...]},
\end{aligned}
\label{eq:Ublack2}
\ee
where $\tilde{B}_b=\sigma^{[0...]z000[0...]}$. This operator has support on one of the four qubits of the corresponding plaquette, $b$, the same qubit that supports the $\sigma^y$ operators in the $C_4$ rotations. These qubits are the ones positioned at the heads of blue arrows in Fig.~\ref{fig:toricmap}(c) to (f). The orientation of unitary parts, and hence these arrows, vary depending on the location of plaquette, $b$, on the lattice. The rule for an arbitrarily sized $d\times d$ lattice with the $X_L$ operators positioned along the central row and column and the $Z_L$ operators along the top row and left column of the lattice, as depicted in Fig.~\ref{fig:toricmap}, goes as follows. The blue arrows of plaquettes in the top right quarter of the lattice point towards the bottom left, those in the bottom left and right quarters point towards the top right and left, respectively, and those in the top left in general point towards the bottom right. There are two exceptions to this rule. One of which is the top left plaquette, which will be mapped to a symmetry operator with support over all other $\tilde{B}_b$ plaquettes, labelled by underlined blue ``$(Z)$"'s in Fig.~\ref{fig:toricmap}(c) to (f). Hence, it does not have a unitary part, $U$, corresponding to it. The second exception is all other plaquettes in the top left quarter of the lattice that run along the diagonal line of black plaquettes from the top left to the bottom right of the lattice. All arrows along this diagonal point towards the top left of the lattice. The orientation of all arrows for the $6\times6$ toric code is shown in Fig.~\ref{fig:toriclarge}.

These $M/2-1$ unitary parts act in a certain order. No unitary part may act before the unitary corresponding to the plaquette their arrow points at. So as can be seen from Fig.~\ref{fig:toricmap} the first plaquette is the one whose arrow points towards the top left plaquette, as this has no unitary part of its own. This ordering ensures that the effect of each part on all other black plaquettes that are yet to be transformed is trivial. However, the effect of each of these parts on the top left plaquette is non-trivial. This is mapped to a symmetry operator, with a $\sigma^z$ supported at each qubit supporting a $\tilde{B}_b$. This non-trivial effect is marked in Fig.~\ref{fig:toricmap}(b) to (f), by the position of underlined blue ``$(Z)$"'s. Similarly to the surface code mapping, the effect of these unitaries on the white plaquettes are also non-trivial. Fig.~\ref{fig:toricmap}(b) to (d) shows that some $\sigma^x$ operators on white plaquettes are mapped to $\sigma^z$ operators by the first $M/2-1$ unitaries. These $\sigma^x$ operators are those with support on the qubits pointed at by the blue arrows.

The next $M/2-1$ unitary parts each act on a white plaquette stabilizer mapping them to single $\sigma^z$ operators. The form of the $U$'s that perform this mapping vary depending on the effect of the $U$'s corresponding to the black plaquettes. The mapping $(U_{M/2-1},...,U_1)$ acts trivially on the two plaquette stabilizers in the bottom left and top right corner of the lattice, as is shown in Fig.~\ref{fig:toricmap}(c), $(U_{N/2}...U_1)B_w(U_1^{\dagger}...U_{N/2}^{\dagger})=B_w=\sigma^{[0...]xxxx[0...]}$. For these types of plaquettes we use the rotation,
\be
B_w\xrightarrow{R_{C_4}(-\sigma^{[0...]xyxx[0...]})}\sigma^{[0...]0z00[0...]}
\ee
All others are acted on non-trivially, such as the leftmost plaquette on the top row, $B_w$, in Fig.~\ref{fig:toricmap}(c), $(U_{M/2-1}...U_1)B_w(U_1^{\dagger}...U_{M/2-1}^{\dagger})=\bar{B}_w=\sigma^{[0...]xxzz[0...]}$, where $\bar{B}_w$ labels the intermediate form of the operator. For an operator of this form we use the rotation,
\be
\bar{B}_w\xrightarrow{R_{C_4}(-\sigma^{[0...]yxzz[0...]})}\sigma^{[0...]z000[0...]},
\ee
where the Pauli operators in the $R_{C_4}$ rotation are equal to those in the operator we are mapping from, $\bar{B}_w$, except at the top left qubit where we replace the $\sigma^x$ with a $-\sigma^y$. The operator on the qubit that will support the final $\tilde{B}_w$ operator will always be unaffected by any previous $U$'s by construction, thus will remain a $\sigma^x$. These qubits are the ones positioned at the heads of orange arrows in Fig.~\ref{fig:toricmap}(e) and (f). The orientation of the second $M/2-1$ unitaries, hence the orange arrows are fixed in a similar way to the first $M/2-1$. The rule for an arbitrarily sized $d\times d$ lattice goes as follows. The orange arrows of plaquettes in the top left and right quarters of the lattice point towards the top left and right, respectively, those in the bottom right quarters point towards the bottom right, and those in the bottom left in general point towards the bottom left. There are again two exceptions to this rule. One of which is the top rightmost plaquette in the bottom left quarter, which will be mapped to a symmetry operator with support over all other $\tilde{B}_w$ plaquettes, labelled by underlined orange ``$(X)$"'s in Fig.~\ref{fig:toricmap}(e) and (f), and thus does not have a unitary part corresponding to it. The second is all other plaquettes in the bottom left quarter of the lattice which run along the diagonal line of white plaquettes from the top right to the bottom left of the lattice. All other arrows along this diagonal point towards the top right of the lattice. The orientation of all arrows for the $6\times6$ toric code is shown in Fig.~\ref{fig:toriclarge}.

As with the first $M/2-1$ unitary parts, the second $M/2-1$ act in a particular order. No unitary part may act before the unitary corresponding to the plaquette their arrow points at. As can be seen from Fig.~\ref{fig:toricmap} the first plaquette must be the one whose arrow points towards the top right plaquette of the bottom left quarter of the lattice, as this has no unitary part of its own. This ordering ensures that the effect of each part on all other white plaquettes that are yet to be transformed is trivial.

Once we have mapped all plaquette operators to single $\sigma^z$ operators or symmetry operators, we transform the logical operators with the four remaining unitaries, $U_{M-1},...,U_{M+2}$. The logical operators, $X_L^{(1)}$ and $X_L^{(2)}$, are mapped by all previous unitaries to strings of $\sigma^z$ operators along the qubits they originally had support on with a $\sigma^x$ on the qubits that intersect with $Z_L^{(1)}$ and $Z_L^{(2)}$, respectively. While $Z_L^{(1)}$ and $Z_L^{(2)}$ are acted on trivially by all previous unitaries. We label these intermediate forms of the operators as $\bar{X}_L^{(1)}=\sigma^{[0...]xzz...[0...]}$, $\bar{X}_L^{(2)}=\sigma^{[0...]xzz...[0...]}$, $\bar{Z}_L^{(1)}=\sigma^{[0...]zzz...[0...]}$ and $\bar{Z}_L^{(2)}=\sigma^{[0...]zzz...[0...]}$. They are shown in Fig.~\ref{fig:toricmap}(e), along with the form of the unitaries $U_{M-1},...,U_{M+2}$, for a $4\times4$ lattice. These act as,
\be
\begin{aligned}
	\bar{X}_L^{(1)}&\xrightarrow{R_{C_4}(\sigma^{[0...]zzz...[0...]})}\sigma^{[0...]y00...[0...]}\xrightarrow{R_{C_4}(-\sigma^{[0...]z00...[0...]})}\sigma^{[0...]x00...[0...]},\\
	\bar{Z}_L^{(1)}&\xrightarrow{R_{C_4}(\sigma^{[0...]xzz...[0...]})}-\sigma^{[0...]y00...[0...]}\xrightarrow{R_{C_4}(-\sigma^{[0...]x00...[0...]})}\sigma^{[0...]z00...[0...]},
\end{aligned}
\ee
where the operators $\bar{X}_L^{(2)}$ and $\bar{Z}_L^{(2)}$ are transformed in a similar way. Thus $X_L^{(1)}$, $Z_L^{(1)}$, $X_L^{(2)}$ and $Z_L^{(2)}$ are mapped to $\tilde{X}_L^{(1)}=\sigma^x_j$, $\tilde{Z}_L^{(1)}=\sigma^z_j$, $\tilde{X}_L^{(2)}=\sigma^x_k$ and $\tilde{Z}_L^{(2)}=\sigma^z_k$. The unitaries $U_{M-1},...,U_{M+2}$ act trivially on all previously obtained $\tilde{B}_b$ and $\tilde{B}_w$ operators.

\subsection{The Fermion Models}

Let us now focus on the properties of the models, $H_{\text{FS}}$ and $H_{\text{FT}}$, that result from the transformations, $\mathcal{U}_S$ and $\mathcal{U}_T$, of the surface and toric code, respectively. We will start with the surface code. The plaquette stabilizers are mapped to $\sigma^z$ operators on free spins (qubits), which are equivalent to free fermion modes. The excitations of plaquettes can now be encoded by the occupation of isolated dynamic fermion modes subject to a local chemical potential that encode the corresponding increase of the energy by $2J$ test when the plaquettes are populated. As a result the Hamiltonian of the transformed model is,
\be
H_{\text{FS}}=-J\sum_{p}\tilde{B}_p,
\label{eq:Hamsurface}
\ee
where $\tilde{B}_p=\sigma^z=1-2a^{\dagger}a$ \cite{NielsenFermiSpinNotes}, and $a^{\dagger}$ and $a$ are fermionic creation and annihilation operators, respectively.

Applying $\mathcal{U}_T$ to a state of the surface code, $\ket{\psi}_{\text{S}}$, gives,
\be
\begin{aligned}
	\mathcal{U}\ket{\psi}_{\text{S}}=&\tilde{\ket{\psi}}_{\text{S}}.
\end{aligned}
\ee
In general for a distance $d$ code, any given state, $\tilde{\ket{\psi}}_{\text{S}}$, has $d^2-1$ dynamic modes, each corresponding to a plaquette of the surface code and one zero mode, which supports the transformed logical operators, $\tilde{X}_L$ and $\tilde{Z}_L$. The ground state, $\ket{\psi_g}_{\text{S}}$, of the original surface code, $H_{\text{SC}}$, is stabilized by all plaquette operators, i.e. $B_p\ket{\psi_g}_{\text{S}}=\ket{\psi_g}_{\text{S}}$ for all plaquettes, $p$. This relationship is preserved by the transformation, $\mathcal{U}_T$. Therefore, $\tilde{B}_p\tilde{\ket{\psi_g}}_{\text{S}}=\tilde{\ket{\psi_g}}_{\text{S}}$ for all $p$, implying the ground state of this model is a collection of $d^2-1$ empty free fermion modes, with a degeneracy of $2$ encoded by the logical zero mode. Occupied dynamic modes indicate the positions of local anyonic excitations in this model.

The transformed $d\times d$ toric code, $H_{\text{FT}}$, has many of the same properties. The Hamiltonian is,
\be
H_{\text{FT}}=-J(\sum_{b\setminus b_1}\tilde{B}_b +\sum_{w\setminus w_1}\tilde{B}_w +\prod_{b\setminus b_1}\tilde{B}_b +\prod_{w\setminus w_1}\tilde{B}_w),
\label{eq:Hamtoric}
\ee
where $b\setminus b_1$ and $w\setminus w_1$ are the sets of all black and white plaquettes, respectively, minus the plaquettes, $b_1$ and $w_1$, that become fermionic symmetry operators over all other transformed plaquettes of the same colour. These symmetry operators are $\tilde{S}_{b_1}=\prod_{b\setminus b_1}\tilde{B}_b$ and $\tilde{S}_{w_1}=\prod_{w\setminus w_1}\tilde{B}_w$ in \rf{eq:Hamtoric}. The other $d^2-2$ transformed plaquette stabilizers have the same form as those in \rf{eq:Hamsurface}, $\tilde{B}_p=\sigma^z=1-2a^{\dagger}a$.

Any transformed state has $d^2-2$ dynamic modes, each corresponding to a plaquette of the toric code and two zero modes, which support the four transformed logical operators. The ground state, $\ket{\psi_g}_{\text{T}}$, of the toric code is stabilized by all plaquette operators, i.e. $B_p\ket{\psi_g}_{\text{T}}=\ket{\psi_g}_{\text{T}}$ for all $p$. Hence the transformed ground state, $\tilde{\ket{\psi_g}}_{\text{T}}$, is stabilized by all transformed plaquette stabilizers, including the symmetry operators, $\tilde{B}_p\tilde{\ket{\psi_g}}_{\text{T}}=\tilde{S}_{b_1}\tilde{\ket{\psi_g}}_{\text{T}}=\tilde{S}_{w_1}\tilde{\ket{\psi_g}}_{\text{T}}=\tilde{\ket{\psi_g}}_{\text{T}}$ for all $p$, implying the ground state is a collection of $d^2-2$ empty free fermion modes, with a degeneracy of $4$ encoded by the two logical zero modes. Occupied dynamic modes indicate the positions of local anyonic excitations in this model. A single occupied mode, $b$, would result in an increase in energy due to the violated stabilizer operator $\tilde{B}_b$, and symmetry operator, $\tilde{S}_{b_1}$. This reflects the fact that excitations are created in pairs at the ends of string operators in $H_{\text{TC}}$, with one end of the string being at plaquette $b$ and one at $b_1$. The symmetry operators, $\tilde{S}_{b_1}$ and $\tilde{S}_{w_1}$, restrict excitations in $H_{\text{FT}}$ to be created in pairs. There is a more detailed discussion of the excitations of $H_{\text{FS}}$ and $H_{\text{FT}}$ in the next section.


\section{Encoding Anyonic Statistics in Free Fermions}

Previous sections have shown that the surface code is unitarily equivalent to a free fermion model, as individual spins are equivalent to free fermion modes. Hence, these models should have equivalent physical properties. Never the less, operators on single free fermions cannot account for the anyonic statistics supported by the surface code. The exotic statistics of its excitations arises due to the commutation and anti-commutation relations of the $\sigma^x$ and $\sigma^z$'s the string operators are built from. In this section we show how these relations are preserved by the unitary transformation, $\mathcal{U}_S$, and how they are encoded in the action of operators on the dynamic and logical modes of the system.

The string operators of the surface code, $O_C^X$ and $O_C^Z$, presented in \rf{eq:stringint}, are a product of $\sigma^x_j$ or $\sigma^z_j$ operators, respectively along the path $C$, producing local excitations at their endpoints. Crossings of these strings give rise to the anyonic statistics through the Pauli commutation relations. These string operators transform as follows,
\be
\begin{aligned}
	\mathcal{U}_SO_C^x\mathcal{U}_S^{\dagger}=\tilde{O}_{\tilde{C}_S}^x, \;\;\;\;\; \mathcal{U}_SO_C^z\mathcal{U}_S^{\dagger}=\tilde{O}_{\tilde{C}_S}^z,
\end{aligned}
\label{eq:stringfreesurface}
\ee
where $\tilde{O}_{\tilde{C}_S}^x$ and $\tilde{O}_{\tilde{C}_S}^z$ are string operators acting on the dynamical and logical fermion modes along the path $\tilde{C}_S$ in $H_{\text{FS}}$.

All commutation relations of operators are preserved by $\mathcal{U}_S$. If $O_C^X$ creates an excitation at plaquette $b$, then $\{O_C^X,B_b\}=0=\{\tilde{O}_{\tilde{C}_S}^X,\tilde{B}_b\}$. If not, $[O_C^X,B_b]=0=[\tilde{O}_{\tilde{C}_S}^X,\tilde{B}_b]$. Hence the endpoints of $\tilde{C}_S$ are the transformed versions of the plaquettes, which were the endpoints of $C$. Paths between endpoints of string operators may change, but the endpoints are fixed at the transformed versions of the plaquettes. Hence, the paths remain homotopically equivalent to those of the untransformed operators. The commutation relations of operators with each other are also preserved, by the mapping $\mathcal{U}_S$. Crossings of these strings may appear in the dynamic or logical modes. Therefore, the anyonic statistics of excitations of the surface code are encoded in the free model by a mix of the dynamic and logical modes.

It is more instructive to look at how string operators, $\tilde{O}_{\tilde{C}_S}$, in the free model, $H_{\text{FS}}$, are mapped under the inverse unitary transformation, $\mathcal{U}_S^{\dagger}$, to string operators, $O_C$, in the surface code, $H_{\text{SC}}$. A $\sigma^x$ operator on a single spin (or $a^{\dagger}+a$ on a single mode) in the free model transforms to a string operator with one end point at the plaquette, $p$, corresponding to that spin (or mode) and one at a boundary not associated with a logical degree of freedom (the top boundary if $p$ is black and the right if $p$ is white). This has to be the case as it is the only type of operator that anti-commutes with just one plaquette. This also suggests why there could not exist a unitary transformation from the toric code with periodic boundary conditions to decoupled free fermions (without the symmetry operators in \rf{eq:Hamtoric}). If each plaquette in the toric code were mapped to a fermion mode in the free model, any operator creating a single fermion population would be mapped to one creating a single plaquette excitation in the toric code. However, all excitations in the toric code must be created in pairs, as dictated by its periodic boundary conditions. In other words the boundary conditions of the surface code are what facilitate such a mapping.

A string operator with end points on any two plaquettes of the same colour in the surface code may be obtained by mapping from a product of two $\sigma^x$'s at the spins (or two $a^{\dagger}+a$ operators at the modes) corresponding to those plaquettes in $H_{\text{FS}}$. The string operators, $\tilde{O}_{\tilde{C}_S}^x$ or $\tilde{O}_{\tilde{C}_S}^z$, that will map to string operators, $O_C^x$ or $O_C^z$, creating logical excitations (i.e. those with end points at the bottom and left boundaries, respectively), contain $\tilde{X}_L$ and $\tilde{Z}_L$, respectively. Any other string operator, $O_C^x$ or $O_C^z$, with the same end points and effect on the logical qubit as those already mentioned may be obtained by including some combination of $\sigma^z$'s in the operators $\tilde{O}_{\tilde{C}_S}^x$ or $\tilde{O}_{\tilde{C}_S}^z$. These $\sigma^z$'s alter the string operator's path by applying stabilizer operations, thus including a loop around the corresponding plaquette to the path, $C$.

The string operators in the toric code are mapped via the unitary transformation, $\mathcal{U}_T$, to string operators in a system of fermion modes coupled to two fermionic parity constraints, in a similar way to those in the surface code,
\be
\begin{aligned}
	\mathcal{U}_TO_C^x\mathcal{U}_T^{\dagger}=&\tilde{O}_{\tilde{C}_T}^x,\\
	\mathcal{U}_TO_C^z\mathcal{U}_T^{\dagger}=&\tilde{O}_{\tilde{C}_T}^z,
\end{aligned}
\label{eq:stringfreetoric}
\ee
where $\tilde{O}_{\tilde{C}_T}^x$ and $\tilde{O}_{\tilde{C}_T}^z$ are string operators acting on the dynamical and logical fermion modes along the path $\tilde{C}_T$ in $H_{\text{FT}}$.

The commutation relations of operators are preserved by $\mathcal{U}_T$. If $O_C^X$ creates an excitation at a black plaquette $b$, then $\{O_C^X,B_b\}=0=\{\tilde{O}_{\tilde{C}_T}^X,\tilde{B}_b\}$ and $\{\tilde{O}_{\tilde{C}_T}^X,\tilde{S}_{b_1}\}=0$. If $b=b_1$ then we say $\tilde{B}_b=\tilde{S}_{b_1}$. If $O_C^X$ does not create an excitation at any black plaquette then, $[O_C^X,B_b]=0=[\tilde{O}_{\tilde{C}_T}^X,\tilde{B}_b]$ and $[\tilde{O}_{\tilde{C}_T}^X,\tilde{S}_{b_1}]=0$ for all $b$. Hence the endpoints of $\tilde{C}_T$ are the transformed versions of the plaquettes, which were the endpoints of $C$. Paths between endpoints of string operators may change, but the endpoints remain fixed. As in the surface code transformation, the commutation relations of operators with each other are also preserved, by the mapping $\mathcal{U}_T$. Crossings of these strings may appear in the dynamic or logical modes. Therefore, as with the surface code mapping, the anyonic statistics of excitations of the toric code are encoded in the fermionic model by a mix of the dynamic and logical modes.

We now look at how string operators, $\tilde{O}_{\tilde{C}_T}$, in the free model, $H_{\text{FT}}$, are mapped under the inverse unitary transformation, $\mathcal{U}_T^{\dagger}$, to string operators, $O_C$, in the toric code, $H_{\text{TC}}$. A $\sigma^x$ operator on a single spin in the free model transforms to a string operator with one end point at the plaquette, $p$, corresponding to that spin and one at the plaquette that was mapped to the symmetry operator of the same colour as $p$ ($b_1$ if $p$ is black and $w_1$ if $p$ is white). This demonstrates how the symmetry operators ensure excitations are created in pairs in $H_{\text{FT}}$, as they are in the toric code.

A string operator with end points on any two plaquettes of the same colour in the toric code may be obtained by mapping from a product of two $\sigma^x$'s at the spins corresponding to those plaquettes in $H_{\text{FT}}$. The string operators, $\tilde{O}_{\tilde{C}_T}^x$ or $\tilde{O}_{\tilde{C}_T}^z$, that will map to string operators, $O_C^x$ or $O_C^z$, with strings of $\sigma^x$'s around a non-contractible loop of the torus, i.e. those which cross the $Z_L^{(1)}$ and/or $Z_L^{(2)}$ operator, contain $\tilde{X}_L^{(1)}$ and/or $\tilde{X}_L^{(2)}$, respectively. Those mapping to operators with strings of $\sigma^z$'s around a non-contractible loop of the torus, i.e. those which cross the $X_L^{(1)}$ and/or $X_L^{(2)}$ operator, contain $\tilde{Z}_L^{(1)}$ and/or $\tilde{Z}_L^{(2)}$, respectively. Any other string operator, $O_C^x$ or $O_C^z$, with the same end points and effects on the logical qubits as those already mentioned are produced by the same method as those in the surface code, by including some combination of $\sigma^z$'s in the operators $\tilde{O}_{\tilde{C}_T}^x$ or $\tilde{O}_{\tilde{C}_T}^z$. These $\sigma^z$'s alter the string operator's path by including a loop around the corresponding plaquette to the path, $C$.


\section{Conclusions and Outlook}
In this paper we have shown that the surface and toric code are unitarily equivalent to free fermions and free fermions coupled to a fermionic parity constraint, respectively. Moreover, we have presented unitary transformations, $\mathcal{U}_S$ and $\mathcal{U}_T$, that map these codes to their fermionic counterparts. We have given the explicit form of these unitaries and demonstrated how the statistical properties of the surface and toric code anyons map to the localised excitations of the fermionic models. We have shown how the periodic boundary conditions of the toric code introduce the need for interacting fermionic parity operators in the fermion model.

We expect the ability to map the surface code to free fermions, could have a number of applications. We have shown, for example, how the anyonic statistics of the excitations are encoded by the unitary transformation. This has allowed for an intuitive and unique understanding of the origins of these statistics. Moreover, descriptions for the construction and manipulation of free fermion systems are more efficient than current interacting descriptions of the surface code \cite{BravyiMajor}. We believe extending the group of mappings, $\mathcal{U}$, to other topological models in two and higher dimensions could provide valuable insight into the emergence of exotic statistics in these systems \cite{DFPRB}.

\section*{ACKNOWLEDGMENTS}
I would like to thank Jiannis Pachos for all his guidance and valuable ideas. This paper would not have been possible without his support. I would also like to thank Christopher Turner for his useful insights and discussions. This research is supported by the Henry Ellison scholarship.

\bibliographystyle{apsrev}



\end{document}